\documentclass[prb,twocolumn,aps,showpacs,fixfloats]{revtex4}
\usepackage{graphicx}
\usepackage{bm}
\usepackage{amsmath, amsthm, amssymb} 
\usepackage{color}
\usepackage{times}
\usepackage{amsfonts}
\usepackage{multirow}
\begin{document}

\title{Binding Energy and Lifetime of Excitons in Metallic Nanotubes}

\author{L.~Shan}
\affiliation{Department of Physics and Astronomy, University of Utah, Salt Lake
City, Utah 84112, USA}

\author{M.~Agarwal}
\affiliation{Department of Physics and Astronomy, University of Utah, Salt Lake
City, Utah 84112, USA}

\author{E.~G.~Mishchenko}
\affiliation{Department of Physics and Astronomy, University of Utah, Salt Lake City, Utah 84112, USA}

\begin{abstract}
The difficulty of describing excitons in semiconducting single-wall nanotubes analytically lies with the fact that excitons can neither be considered strictly one- nor two-dimensional objects. However, the situation changes in the case of metallic nanotubes where, by virtue of screening from gapless metallic subbands, the radius of the exciton becomes much larger than the radius of the nanotube $R_\text{ex}\gg R$. Taking advantage of this, we develop the theory of excitons in metallic nanotubes, determining that their binding energy is about $0.08 v/R$, in agreement with the existing experimental data. Additionally, because of the presence of the gapless subbands, there are processes where bound excitons are scattered into unbound electron-hole pairs belonging to the gapless subbands. Such processes lead to a finite exciton lifetime and the broadening of its spectral function. We calculate the corresponding decay rate of the excitons.
\end{abstract}


\maketitle

\section{Introduction}
An exciton is a bound state of an electron and a hole formed by their Coulomb attraction--the solid state analogy of a hydrogen atom, but with a larger size and a smaller binding energy. The smaller binding energy is due to the screening of the mutual Coulomb interaction by bound electrons of the medium, described by its dielectric constant. Excitons typically exist in insulators and weakly doped semiconductors. Metallic and strongly doped semiconducting materials disfavor formation of excitons. This is the result of two factors. First, as the conduction band is populated (e.g. by doping), screening by free charges strongly reduces  the magnitude of the electron-hole Coulomb interaction and decreases its binding energy. Second, population of the conduction band reduces the number of quantum states available to accommodate the electron after its (virtual) scattering off the hole, further decreasing the binding energy, to the point where no meaningful bound state may be formed anymore.

This situation changes in quasi-one-dimensional systems, such as metallic nanotubes,\cite{Dresselhaus} where  formation of excitons occurs in subbands {\it  different} from the subbands that are responsible for metallic screening. Such separation occurs as a result of quantization of the circumferential momentum. This makes exciton a well-defined excitation. Excitons in metallic single-walled carbon nanotubes (SWNTs) were experimentally observed in Refs.~[\onlinecite{WCKDSLZHS,MTZRTM}]. Their binding energy was found to be about 50 meV, an order of magnitude smaller than the typical bandgaps  $\Delta$ of semiconducting nanotubes. Theoretically, excitons in metallic SWNTs were first studied by Ando under the effective-mass approximation.\cite{Ando} Later, the binding energy of the exciton was addressed  by first principles calculations\cite{SBBL,DSPL,UA} and also by the density matrix theory,\cite{MMRK} with the latter results being in agreement with experimental measurements. A brief review of excitonic effects in metallic SWNTs was given in Ref.~[\onlinecite{AU}]. However, to our knowledge, no simple analytical description of excitons in metallic nanotubes has been developed. It is the purpose of the present paper to fill this gap.

Let us illustrate the difficulty of describing excitons analytically in semiconducting SWNTs,  stemming from the fact that excitons can neither be considered one-dimensional (1D) nor two-dimensional (2D) excitations. Indeed, consider the lowest energy subbands with the spectrum $\varepsilon (p) =\pm \sqrt{\Delta^2+v^2p^2}$, and expand it near the bottom of the subband,
\begin{equation}
\label{effective mass}
\sqrt{\Delta^2+v^2p^2} \approx \Delta+\frac{p^2}{2\mu}, \hspace{0.4cm} \mu =\frac{\Delta}{v^2}.
\end{equation}
The bandgap is typically, $\Delta \sim  v/R$, where $R$ is the radius of the nanotube. For convenience, we set $\hbar =1$ throughout the paper. (In particular, in the zone-folding tight-binding approximation both the (8,0) and (10,0) zigzag nanotubes have $\Delta = v/3R$).

Because the electron-hole interaction energy is $U(r) =- e^2/r$, the exciton binding energy $E_b$ can be estimated by minimizing,
\begin{equation}
-E_b \approx\text{min}\left[ \frac{v^2p^2}{2\Delta}-\frac{e^2}{r}\right],
\end{equation}
taking into account the uncertainty relation, $r\sim 1/p$. This yields,
\begin{equation}
E_b \sim \left(\frac{e^2}{ v}\right)^2 \Delta.
\end{equation}
Since $e^2/ v \sim 1$, we obtain that $E_b \sim \Delta$ and, consequently,  the exciton radius $R_{ex} \sim  v/E_b \sim R$. Because to consider excitons to be 1D one would need to have $R_{ex} \gg R$, and conversely, to view them as 2D one would require $R_{ex} \ll R$, the semiconducting problem falls instead between the two limits where a numerical analysis is necessary. However, as we are going to see below, the situation changes in the case of metallic nanotubes where--by virtue of screening by the metallic subbands--the exciton binding energy decreases significantly (as already evidenced by the experimental data). As a result, the radius of the exciton increases, $R_{ex} \gg R$, and treatment of the exciton as a quasi-one-dimensional object becomes possible. This is what makes the analytic solution viable.

In this paper, we determine the binding energy and lifetime of an exciton in metallic SWNTs taking into account the screening effects within the random phase approximation (RPA). A problematic feature of the 1D Coulomb problem is the $r^{-1}$ singularity in the potential energy, which is not integrable (unlike in the three-dimensional situation). This feature was first addressed by Loudon \cite{Loudon} and later further extensively studied  by others.\cite{Andrew1,HR,Andrew2,GZ,Moss} Loudon introduced the truncated Coulomb interaction $e^2/(|x|+a_0)$ with a positive constant $a_0$ to ensure that Coulomb potential is regular at small distances $x\rightarrow0$. In our problem, the nanotube radius $R$ appears naturally and no other cutoff is needed. Using a variational ansatz with a Gaussian trial function, we show below that the binding energy is (the value of $\Delta$ corresponds to an armchair metallic nanotube)
\begin{equation}
\label{binding energy intro}
E_b\approx0.08 \, \Delta, \hspace{0.4cm} \Delta =\frac{v}{R}.
\end{equation}

The $1/R$ dependence of the binding energy exhibits a good agreement with the results obtained in Ref.~[\onlinecite{MMRK}] with the density matrix theory, see Table~I. In contrast, {\it ab initio} methods yield only rather crude estimates of the binding energy.\cite{WCKDSLZHS,SBBL,DSPL} This is because computational limitations do not permit calculations for large nanotubes. In addition, the technique of photoluminescence (PL) spectroscopy cannot be applied to obtain binding energies of metallic nanotubes, which makes it challenging to determine the binding energy of excitons in metallic SWNTs. This is because the exciton is likely to undergo a nonradiative decay to the nearby linear subbands instead of electron-hole recombination process with a photon released. Nonetheless, the exciton binding energies experimentally observed\cite{WCKDSLZHS,MTZRTM} have the same order of magnitude as predicted by our calculations.

\begin{table}[h]
\centering
\renewcommand\arraystretch{2}
\caption{Analytically and numerically calculated (or experimentally measured) binding energies of the excitons in different metallic carbon nanotubes (M-NTs). The values in parenthesis are extrapolated from the binding energy for the $(13,1)$ nanotube obtained in Ref.~[\onlinecite{MMRK}] and reported in it $E_b \propto 1/R$ dependence of the binding energy on the radius of a metallic nanotube.}
\begin{tabular}{|c|c|c|c|c|}
\hline
\multirow{2}{*}{M-NTs $(n,m)$} & \multicolumn{4}{|c|}{Binding energies $E_b$ (meV)}\\
\cline{2-5}
& This paper &  Density matrix & {\it ab initio} & Exper. \\
\hline
(12,0) & 86  & (88) & $\sim 50$[\onlinecite{DSPL}] &-- \\
\hline
(13,1) & 76  & 78[\onlinecite{MMRK}]& -- &$\sim 50$[\onlinecite{MTZRTM}] \\
\hline
(10,10) & 60  & (61) & $\sim 50$[\onlinecite{DSPL}]&-- \\
\hline
(21,21) & 28  & (29) & -- & $\sim 50$[\onlinecite{WCKDSLZHS} ]\\
\hline
\end{tabular}
\end{table}

\begin{figure}
\resizebox{.4\textwidth}{!}{\includegraphics{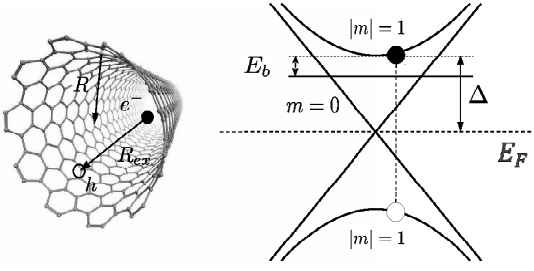}}
\caption{\label{fig1} Exciton state emerges due to the final state interaction of the electron and hole with energy below the parabolic subband. }
\end{figure}

Furthermore, we explore the stability of excitons in metallic SWNTs, the question that is of fundamental interest but which remains unexplored in the existing theoretical works. The mechanism of a finite lifetime  of the exciton can be illustrated by Fig.~\ref{fig1}. The $m$-th subband, corresponding to the integer angular quantum number $m$, has the energy $\varepsilon_m(p)=\pm v\sqrt{m^2/R^2+p^2}$. Exciton bound states are formed between gapped, $m\ne 0$, subbands, and the lowest $m=\pm 1$ exciton is indicated by a dashed line in Fig.~\ref{fig1}. This energy {\it overlaps} with the gapless $m=0$ subbands. Accordingly, an elastic transition of the electron and the hole from gapped to gapless subbands opens up a decay channel for the exciton.

\begin{figure}
\resizebox{.48\textwidth}{!}{\includegraphics{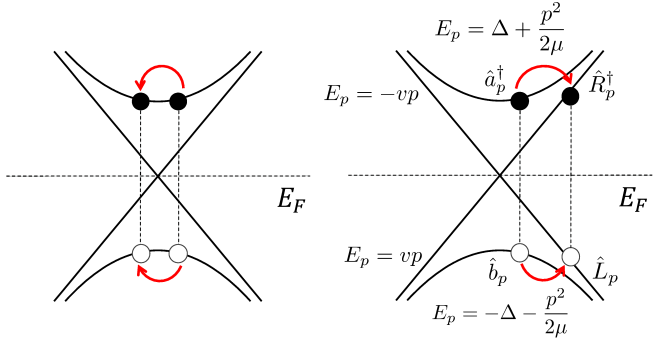}}
\caption{\label{fig2} Scattering processes shown within the massive subband (via $U_{0}$ coupling) and between massive and massless subbands (via $U_{1}$ coupling). }
\end{figure}

To better understand the role of the Coulomb interaction in the formation and decay of the exciton, the following picture is helpful. The exciton is produced by multiple virtual scattering events of the electron and the hole within the gapped subbands. Such transitions are controlled by the Coulomb coupling $V_0$, the subscript indicating that no angular momentum change takes place.  These intrasubband transitions determine the binding energy of the exciton, see Fig.~\ref{fig2}, left panel. The intersubband transitions--shown on the right panel in Fig.~\ref{fig2}--appear as a result of the ``dipolar'' Coulomb interaction $V_1$ and occur with a change $\pm 1$ of the angular momentum of the electron (and with the opposite change of the angular momentum of the hole). We obtain that the ratio of the binding energy to the half-width of the exciton spectral function is
\begin{equation}
\label{ratio E to G}
{\Gamma} \approx 0.015\, \Delta.
\end{equation}
Such a ratio indicates that excitons in metallic SWNTs are well-defined excitations. This is consistent with experimental results obtained through ultrafast luminescence.\cite{KSSMSN} Below we derive our main results, Eqs.~(\ref{binding energy intro}) and (\ref{ratio E to G}).

\section{The Hamiltonian of the system}

The Coulomb  interaction potential between two electrons located on the surface of a nanotube of radius $R$ at $(x_1,\vartheta_1)$ and $(x_2,\vartheta_2)$ is given by,
\begin{equation}
\label{real_space}
V(x,\vartheta) = \frac{e^2}{\sqrt{x^2+4R^2\sin^2(\vartheta/2)}},
\end{equation}
where $x=x_1-x_2$ and $\vartheta=\vartheta_1-\vartheta_2$ are the relative distance along the nanotube axis and the relative angle around the circumference of the nanotube. [In the case of an electron-hole pair, the sign of the interaction (\ref{real_space}) is reversed.] If $R$ becomes large compared with a characteristic distance of the electron and the hole trajectories, the interaction approaches the Coulomb potential on a flat two-dimensional plane. On the other hand, if $R$ becomes small, Eq.~(\ref{real_space}) reduces to the ``1D hydrogen'' problem studied by Loudon. \cite{Loudon} The Coulomb interaction on the cylinder--the problem interpolating between $1D$ and $2D$--was previously considered by Mahan\cite{KCM} and Petersen.\cite{Petersen}

What is different in our case is the need to take into account screening, introduced by the gapless subbands of metallic SWNTs.
This is most conveniently done by transitioning to the momentum space. The Fourier transform of the Coulomb interaction (\ref{real_space}) yields the following expression,\cite{Ando, MAG}
\begin{eqnarray}
\label{q_space}
V_m(q) &=& \int\limits_{-\infty}^{\infty} dx ~e^{-iqx}\int\limits_{-\pi}^{\pi} \frac{d\vartheta}{2\pi} ~e^{-im\vartheta}\frac{e^2}{\sqrt{x^2+4R^2\sin^2(\vartheta/2)}}, \nonumber\\
&=& {2e^2} K_{|m|}(qR) I_{|m|}(qR),
\end{eqnarray}
where $m$ is an integer number, and $I_{m}$ and $K_{m}$ are the modified Bessel functions of the first kind and second kind, respectively.
However, because of the presence of conduction electrons in the gapless subbands, which move around in response to the ``bare'' Coulomb interaction $V_m(q)$, the actual (screened) interaction is modified: $V_m(q) \to U_m(\omega, q)$. This screened interaction disturbs the equilibrium of the system and induces density variations, which within the linear response are proportional to the strength of the interaction, $n_m(\omega,q)=\Pi_m (\omega, q) U_m(\omega, q)$, with the coefficient $\Pi_m (\omega, q)$ referred to as the polarization function (i.e. density-density correlation function) associated with the change $m$ in the angular momentum and the change $q$ in the linear momentum. The central tenet of RPA is the assumption of the mean field, which predicts that the variation of the density $n_m(\omega,q)$ induces the additional potential in the system: $e\varphi (\omega,q) =V_m(q)n_m(\omega,q)=V_m(q) \Pi_m (\omega, q) U_m(\omega, q)$. This additional potential together with the bare potential constitutes the total interaction: $e\phi (\omega,q) +V_m(q)= U_m(\omega, q)$. This gives,
\begin{equation}
U_m(\omega, q) =\frac{V_m(q)}{1-V_m(q) \Pi_m (\omega, q)}.
\end{equation}

The formation of the exciton is mostly facilitated by the strongest interaction  $V_0$, which at low $qR
\ll 1$ becomes logarithmically strong, $V_0(q) \approx -2e^2 \ln{(qR)}$. In contrast, the higher interaction harmonics, $m \ne 0$,  remain constant in this limit, $V_m\to e^2/|m|$.
The screening of the $V_0$ interaction
is determined by the uniform harmonics of the polarization function with no change in the angular momentum, given by,
\begin{equation}
\label{polarizationV_0}
\Pi_0(\omega, q) =\frac{N}{\pi  v} \frac{q^2 v^2}{\omega^2-q^2 v^2 },
\end{equation}
where $N=4$ accounts for two spin directions and the presence of the two Dirac points within the Brillouin zone (see Appendix for details).

A large spatial radius of the exciton $R_{ex}$ makes it sufficient to consider only the static limit of Eq.~(\ref{polarizationV_0}), where $\omega$ is disregarded compared with $qv$. Indeed, frequencies involved are of the order of the exciton binding energy, $\omega \sim E_b = 1/(\mu R^2_{ex})$, where the effective mass $\mu\sim 1/(vR)$, according to Eq.~(\ref{effective mass}). On the other hand, the involved momenta are  $q\sim 1/R_{ex}$. Accordingly, the ratio $\omega^2/q^2v^2 \sim R^2/R^2_{\rm ex} \ll 1$. The screened Coulomb interaction in this static limit assumes the form,
\begin{equation}
\label{U_0}
U_0(q) = \frac{2e^2K_0(qR)I_0(qR)}{1+\alpha K_0(qR)I_0(qR)},~~~\alpha=\frac{2Ne^2}{\pi v} \approx 6.9,
\end{equation}
where we introduced the dimensionless interaction strength $\alpha=2Ne^2/\pi v$ where $v$ is taken to be the same as the velocity of electrons propagating in graphene, $v = 8 \times 10^5$ m/s.

To similarly calculate the screening of the $V_{1}$ interaction, one needs to know the polarization function associated with the $\pm 1$ change of the angular momentum. While the intrasubband value of $\Pi_0(\omega, q)$ at low $\omega$ and $qv$, as seen from Eq.~(\ref{polarizationV_0}), depends on which one of the two quantities tends to zero faster than the other, there is no such ambiguity for the intersubband polarization function $\Pi_1(\omega, q)$: as shown in Appendix, $\Pi_1(0,0)=-1.16N/(\pi v)$. 
This gives for the screened intersubband interaction,
\begin{equation}
\label{U1}
U_1 = \frac{e^2}{1 +1.16\frac{e^2N}{\pi  v}} \approx 0.2 \,e^2.
\end{equation}
In the limit of large wavelengths, $qR\ll 1$, the two interactions approach each other, $U_0(0)\approx U_1$. It should be pointed out, however, that while the $U_1$ interaction remains almost constant for finite but small $q$, the interaction $U_1(q)$ is rather sensitive to $q$, with the derivative $dU_1(q)/dq$ diverging at $q\to 0$.

Note that in the absence of spectrum curvature, the polarization function $\Pi_0(\omega,q)$ of one-dimensional subbands is independent of temperature. It is also unmodified by the electron-electron interactions.\cite{DL}  On the other hand, the polarization constant $\Pi_1(0,0)$, which involves gapped subbands, has a negligible temperature dependence. Indeed, the population of thermally excited electron-hole pairs is $\sim \exp(-\Delta/k_{B}T)$, where $\Delta$ is much larger than the thermal energy at room temperature. For these reasons it is sufficient to consider the screening of the Coulomb interaction at zero temperature. Similarly, small levels of doping, $\mu \ll \Delta$, do not affect the strength of the Coulomb interaction.

Having determined the magnitude of the electron-electron (and, therefore, electron-hole) interaction, we can proceed to calculate the energy and the lifetime of the exciton. The Hamiltonian of the system has the following form:
\begin{align}
\label{Hamiltonian}
\hat H = &\sum_p  \left(\frac{p^2}{2\mu}+\Delta\right) ( \hat a_p^\dagger \hat a_p - \hat b_p^\dagger \hat b_p ) \nonumber\\
 &+  \sum_p vp \,( \hat R^\dagger_p \hat R_p - \hat L^\dagger_p \hat L_p ) \nonumber\\
 &+ \frac{1}{L}\sum_{p,k} U_0(p-k)\hat a^\dagger_k b^\dagger_p  \hat b_k \hat a_p \nonumber\\
 &+\frac{U_1}{2L}\sum_{p,k} (\hat R_k^\dagger b^\dagger_p  \hat L_k \hat a_p+ a^\dagger_p \hat L^\dagger_k  \hat b_p \hat R_k)\nonumber\\
 &+\frac{U_1}{2L}\sum_{p,k} (\hat L_k^\dagger b^\dagger_p \hat R_k \hat a_p  + \hat a^\dagger_p \hat R^\dagger_k  \hat b_p \hat L_k).
\end{align}
Here $\hat a_p$ and $\hat b_p$ are the operators for the particles residing on the upper and lower subbands with $m=1$ (or $m=-1$); the operators $\hat R_p$ and $\hat L_p$ correspond to right and left moving particles of massless ($m=0$) subbands; the $U_0(q)$ interaction describes scattering within the massive subbands whereas $U_1$ coupling describes processes where scattering occurs between massive and massless subbands. Note that of all of the Coulomb interaction terms we have retained only those that are responsible for the formation of the excition with zero total momentum: for example, the $U_0$-term describes scattering of the electron with momentum $p$ ($\hat a_p$) and the hole with momentum $-p$ ($\hat b_p^\dagger$) into a pair of new states with momenta $k$ and $-k$.

The intersubband terms in the Hamiltonian (\ref{Hamiltonian}) have the extra prefactor $1/2$ compared with the intrasubband transitions. The origin of this difference lies in the pseudospin nature of the underlying Hamiltonian of the two-dimensional graphene crystal that forms the nanotube. (The pseudospin arises from the existence of two atomic sublattices in the graphene honeycomb arrangement of carbon atoms.) As a result, two states of the same energy and opposite momenta are orthogonal to each other. More generally, the amplitude of the transition  between the states having momenta ${\bf p}$ and ${\bf p}+{\bf q}$ and the same sign of energy is suppressed by the factor $\cos[(\theta_{\bf p+q} - \theta_{\bf p})/2]$, where $\theta_{\bf p}$ is the angle that the momentum ${\bf p}$ makes with the $x$-axis. When graphene is rolled into a nanotube, the circumferential momenta are quantized, $p_y=m/R$. The gapless states $m=0$ are those that move along the $x$-axis: ${\bf p}=(p,0)$ with $\theta_{\bf p} =0$ or $\pi$. In contrast, in the gapped subbands, ${\bf p}= (p, m/R)$, with the relevant momenta are near the bottom/top of the subbands: $p \ll 1/R$. Accordingly, the relevant states are those that have $\theta_{\bf p} \approx \pm \pi/2$. Correspondingly, each particle transitioning between a gapless state and a gapped state (close to the bottom of the subband) introduces a factor $\cos{(\pi/4})=1/\sqrt{2}$ into the amplitude of the scattering. For the two-particle Coulomb interaction $U_1$, the total additional coefficient is, therefore, $1/2$.

\section{Binding energy and lifetime}

The wave function of the exciton is sought in the form
\begin{equation}
\label{wavefunction}
|\psi\rangle = \sum_q f_q \hat a_q^\dagger \hat b_q |0\rangle + \sum_{q>0} g_q \hat R^\dagger_q \hat L_q |0\rangle +\sum_{q<0} g_q \hat L^\dagger_q \hat R_q|0\rangle,
\end{equation}
where $|0\rangle$ is the ground state of the system where all individual electron states with the positive energy are empty and all states with the negative energy are occupied.
 The function $f_p$ describes the amplitude of the electron-hole pair to be in the gapped states whereas the function $ g_q$ describes the likelihood of the pair to reside in the massless states. Note that in the last two terms we have explicitly taken into account that in the ground state $|0\rangle$ the left-moving states are occupied as long as $q$ is positive whereas the right-moving states are occupied if $q$ is negative. Because the system is symmetric with respect to the symmetry between left- and right-moving states, the function $g_q$ must be symmetric: $g_q =g_{-q}$.

The Schr{\"o}dinger equation $\hat H |0\rangle= E |0\rangle$ separates into two coupled equations for the  functions $f_q$ and $g_q$:
\begin{align}
\label{eqnset0}
&\left( E - 2\Delta - {q^2}/{\mu} \right)f_q = -\frac{1}{L} \sum_p U_0(q-p) f_p -\frac{U_1}{2L} \sum_{p}   g_p, \nonumber\\
& \left(E - 2v|q| \right)g_q = -\frac{U_1}{2L} \sum_{p}f_p.
\end{align}

Excluding $g_q$ from these equations and replacing the sums by the integrals, $\sum_{p} \rightarrow L\int dp/2\pi$, we arrive at the following  integral equation for the function $f_q$,
\begin{align}
\label{eqnset1}
\left( E - 2\Delta - {q^2}/{\mu} \right)f_q =& \int\limits_{-\infty}^{\infty} \frac{dp}{2\pi} f_p\Bigl[ -U_0(p-q) , \nonumber\\
&+ \frac{U_1^2}{4} \int\limits_{-\infty}^{\infty} \frac{dp'}{2\pi} \frac{1}{E-2|p'|v+i\eta} \Bigr].
\end{align}

The singularities at $E=2|p'|v$ lead to the imaginary part in the energy $E-i\Gamma/2 $ of the exciton.
Because the resulting imaginary part is small compared with the bandgap, $\Gamma \ll \Delta$, it is sufficient to keep the infinitesimal $\eta$ in the denominator and utilize the Sokhotski  identity, $\text{Im}\,1/(E-2|p'|v+i\eta) = -i\pi	 \delta(E-2|p'|v)$,  for the calculation of the integral's imaginary part. In contrast, the real part, which arises from the principal value of the integral, can be ignored. Although the real part appears to diverge logarithmically, such divergence is the artefact of our assumption that $U_1$ is constant. This approximation, in any case, fails for transferred momenta of the order $1/R$, which should, therefore, be used as the upper cut-off for the logarithmic integral. Finally, because the energy involved is large, $E\sim 2\Delta=2v/R$ (rather than the small binding energy $E_b$), the logarithm is of the order $1$ and the real part of the last term in the brackets in Eq.~(\ref{eqnset1}) merely adds a contribution $\sim U_{1}^2/8\pi v$. This second-order correction is small compared to the main contribution from the $U_{0}$ term and may be ignored. Hence, we obtain from Eq.~(\ref{eqnset1}),
\begin{equation}
\label{lifetime}
\left( E - 2\Delta - \frac{q^2}{\mu} \right)f_q = -\int\limits_{-\infty}^{\infty} \frac{dp}{2\pi} \Bigl[ U_0(p-q) + i\frac{U_1^2}{8v} \Bigr] f_p.
\end{equation}
Below, the binding energy and the lifetime of the exciton,
\begin{equation}
E=2\Delta-E_b -i\Gamma/2,
\end{equation}
are determined from the ground-state eigenvalue $E$ of this equation.

\subsection{The shallow potential approximation}

The fastest way to {\it estimate} the exciton binding energy is by utilizing the well-known in quantum mechanics shallow well approximation which in the momentum space amounts to replacing the interaction $U_0(q)$ with its zero-momentum value $U_0\equiv U_0(0)$. Equation (\ref{lifetime}) then becomes exactly solvable and yields,
\begin{equation}
E_b+i\Gamma/2 =  \frac{\mu}{4}\Bigl[ U_0 + i\frac{U_1^2}{8v} \Bigr]^2 \approx \frac{\mu}{4}\Bigl[ U_0^2 + i\frac{U_0U_1^2}{4v}\Bigr].
\end{equation}
The effective mass, cf. Eq.~(\ref{effective mass}), can now be expressed via the bandgap, $\mu =\Delta/v^2$ and the interaction constants $U_0 \approx \pi v /4$ and $U_1 \approx 0.2\,e^2$, given by Eqs.~(\ref{U_0}) and (\ref{U1}).  This gives,
\begin{equation}
\label{Ebshallow}
E_b = \frac{\pi^2}{64} \Delta,\hspace{0.4cm} \Gamma/E_b \approx 0.19.
\end{equation}

It is clear, however, that the value $E_b=0.15 \Delta$ that follows from Eq.~(\ref{Ebshallow}), significantly {\it overestimates} the binding energy of the exciton: as evidenced by Eq.~(\ref{U_0}), the function $U_0(q)$ is irregular at $q\to 0$ where it has an infinite derivative. For this reason, the integrand in Eq.~(\ref{lifetime}) in fact contributes much less to the integral where  $p$ departs from $q$ than predicted by the shallow potential approximation. To make a better approximation, we are now going to utilize the variational approach.

\subsection{The variational solution}

To apply the variational approach to  Eq.~(\ref{lifetime}) we choose the Gaussian trial function,
\begin{equation}
\label{trial_x}
f (x) = \left(\frac{\beta}{\pi}\right)^{1/4}e^{-\beta x^2/2},
\end{equation}
or, equivalently, in the momentum space, $f_p = ({4\pi}/{\beta})^{1/4}e^{-{p^2}/{2\beta}}$.
According to Eq.~(\ref{lifetime}), the ground state energy of the exciton is the sum of the kinetic energy,
\begin{equation}
\overline{T}= \int\limits^\infty_{-\infty} \frac{dp}{2\pi}  \frac{p^2}{\mu} f_p^2 = \frac{\beta}{2\mu},
\end{equation}
and the expectation value of the potential energy. The latter is  a complex quantity: $U_0(x) -i\frac{U_1^2}{8v} \delta(x) $.
The imaginary part determines the exciton lifetime,
\begin{equation}
\label{lifetime variational}
\Gamma/2= \frac{U_1^2}{8v} [ f(0)]^2 = \frac{U_1^2}{8v}\sqrt{\frac{\beta}{\pi}},
\end{equation}
while the real part yields, with the help of the momentum representation, the following integral:
\begin{equation}
\label{average_p}
\overline{U_0} =   \int\limits^\infty_{-\infty} dx~U_0(x) [f(x)]^2 = \int\limits^\infty_{-\infty} \frac{dp}{2\pi}~U_0(p) e^{-{p^2}/{4\beta}}.
\end{equation}
The binding energy should be found by minimizing the sum $\overline{T}-\overline{U_0}$ with respect to $\beta$. The integral in $\overline{U_0}$ cannot be calculated exactly, but can be approximated rather accurately. First, it is convenient to utilize the dimensionless variables $s=pR$ and $t=\sqrt\beta R$ to recast the average potential energy (\ref{U_0}) in the form:
\begin{equation}
\overline U_0 = \frac{2e^2}{\pi R} \int\limits^\infty_0 ds~e^{-s^2/4t^2} \frac{K_0(s)I_0(s)}{1 + \alpha K_0(s)I_0(s)},
\end{equation}
Next, we anticipate that for the large-radius excitons the small values of $t<1$ (and hence the small values of $s$) are relevant  (This expectation that is supported by the final result).  The function $I_0(s)\approx 1$, whereas the Macdonald function has a logarithmic singularity, $K_0(s)\approx \ln{(2/s)}-\gamma$, where $\gamma =0.577$ is the Euler constant.  Finally, we notice that the logarithm depends on its argument $s$ rather weakly, in comparison with the exponential $e^{-s^2/4t^2}$, and hence can be approximated as a constant within the relevant range of the $s$-integration,
\begin{equation}
\label{integral approximate}
\int\limits^\infty_0 ds~e^{-s^2/4t^2} \frac{K_0(s)I_0(s)}{1 + \alpha K_0(s)I_0(s)} \approx   \frac{\sqrt{\pi} \, t \ln{(C/t)}}{1 + \alpha \ln{(C/t)}},
\end{equation}
with some fitting parameter $C$. Numerical calculation demonstrates that the value $C=3/2$ provides an excellent fit between the exact numerical integration of the left-hand side of Eq.~(\ref{integral approximate}) and its right-hand side for the value $\alpha=6.9$ stated in Eq.~(\ref{U_0}), which corresponds to $N=4$ gapless modes.

Accordingly, we arrive at the following value of the binding energy, $-E_b= \overline{T}-\overline{U_0}$,
\begin{align}
\label{binding energy function}
{E}_b = &-\frac{t^2}{2\mu R^2} + \frac{2e^2}{\sqrt\pi R} \frac{t \ln{(C/t)}}{1 + \alpha \ln{(C/t)}}\nonumber\\
&=\Delta \left[- \frac{t^2}{2} + \frac{\sqrt{\pi}}{N} \frac{\alpha t \ln{(C/t)}}{1 + \alpha \ln{(C/t)}}\right],
\end{align}
where in the last line we used the definition of the coupling constant $\alpha$ from Eq.~(\ref{U_0}) and also replaced the effective mass $\mu=\Delta/v^2$ in terms of the bandgap $\Delta =v/R$.

The binding energy (\ref{binding energy function}) has a maximum at $t=t_0=0.36$, where the value $E_b$ stated in Eq.~(\ref{binding energy intro}) is reached.  This binding energy agrees well with the experimental measurements {\cite{WCKDSLZHS}. In turn, the obtained result justifies the approximations of a large exciton radius. Indeed, according to the wave function (\ref{trial_x}), the radius of the exciton is $R_{ex} \sim \sqrt{2/\beta}=\sqrt{2}R/t_0 \approx 4 R$.

The exciton radius increases when the Coulomb interaction is further reduced if the nanotube is located on the surface of a dielectric substrate.  Correspondingly, the dimensionless interaction strength will be modified as, $\alpha = 4 N e^2/\pi v (\kappa+1)$, where $\kappa$ is the dielectric constant of the substrate. The binding energy can still be calculated using Eq.~(\ref{binding energy function}). Table II shows the calculated binding energies for several substrates.

The decay of the exciton into the linear subbands was first studied numerically in Ref.~[\onlinecite{UA}] which concluded that such decay processes lead to a negligible broadening. Our result indicate otherwise and are also in a good agreement with experiments reported in Ref.~[\onlinecite{KSSMSN}].


\begin{table}[h]
\centering
\renewcommand\arraystretch{2}
\caption{The binding energies of the excitons in $(21,21)$ armchair nanotubes for different dielectric constant $\kappa$.}
\begin{tabular}{|c|c|c|}
\hline
Substrates & Dielectric constant $\kappa$& Binding energy $E_b$ (meV) \\
\hline
$\text{SiO}_2$ & 2.5 & 22.4 \\
\hline
SiC & 3.75 & 18.7\\
\hline
$\text{Si/SiO}_2$ & 4.4 & 17.9\\
\hline
GaAs & 7 & 13.8\\
\hline
\end{tabular}
\end{table}

The value of the inverse lifetime now follows from Eq.~(\ref{lifetime variational}):
 \begin{equation}
 \label{lifetime final result}
 \Gamma = \frac{\Delta t_0}{4\sqrt{\pi}}\frac{U^2_1}{ v^2},
 \end{equation}
 whose numerical value yields Eq.~(\ref{ratio E to G}).  The exciton acquires a significant broadening, but nonetheless remains a well-defined excitation.  Interestingly, this ratio is numerically very close to the value predicted  by the shallow potential  approximation.

\section{Summary and conclusions}

Because of the presence of the gapless subbands, excitons in metallic carbon nanotubes acquire unique features that distinguish them from excitons in other solid state systems.
First, the quasi-one-dimensional nature of nanotubes makes screening by conduction electrons less effective than in conventional metals. As a result, the electron-hole interaction remains significant enough to ensure the formation of a bound pair. Second, the separation  (in the momentum space) of gapless states ($m=0$) from the subbands where the exciton is formed ($|m|=1$) and the fact that the latter subbands are fully gapped allow the electron and the hole to explore fully the gapped subbands, unlike what happens in a conventional doped semiconductor where filling of the conduction band quickly depletes the number of available electron states. Third, the screening by the gapless states is nonetheless significant enough so that the radius of the exciton is greater than the nanotube radius with the binding energy of the order of $0.1\Delta$. This allows one to treat excitons as quasi-one-dimensional objects, unlike excitons in semiconducting nanotubes which are neither one-dimensional nor two-dimensional objects. Fourth, the presence of the gapless subbands opens up a channel for exciton attenuation where the electron and hole can scatter off each other into the gapless states. The presence of this channel leads to a considerable broadening of the exciton but not so significant as to smear it away completely.

This work was supported by the DOE, Office of Basic Energy Sciences, through Grant No.~DE-FG02-06ER46313.

\appendix
\section{Polarization function of a metallic nanotube}

The screening of Coulomb interaction in a nanotube is determined by its polarization function $\Pi_m(i\omega, q)$ which in the zone-folding approximation
can be obtained from the polarization function of the underlying two-dimensional graphene crystal. The low-energy excitations in graphene are described by the Dirac Hamiltonian, $\hat H =  v {\bm \sigma}\cdot {\bm p}$, where ${\bm \sigma}$ is the Pauli matrix acting in the pseudospin space of the two triangular sublattices of carbon atoms. The polarization function,
\begin{equation}
\Pi(i\omega,{\bf q}) = NT \text{Tr} \sum_{i\epsilon} \int \frac{d^2p}{(2\pi)^2} \hat G^{(0)}(i\epsilon+i\omega,{\bf p+q})\hat G^{(0)}(i\epsilon,{\bf p}),
\end{equation}
relates to the product of two Green's function of $\pi$-electrons consisting of the contributions from both the upper cone ($\beta=1$) and lower cone ($\beta=-1$),
\begin{equation}
G^{(0)}(i\epsilon,{\bf p}) = \frac{1}{i\epsilon - v{\bf p}\cdot \hat {\bm \sigma}} = \frac{1}{2}\sum_{\beta=\pm1}\frac{1+\beta \hat {\sigma_p}}{i\epsilon - \beta vp},
\end{equation}
where $\hat {\sigma_p} = \hat {\bm \sigma}\cdot {\bf p}/p$ is the projection of the pseudospin Pauli matrix onto the direction of electron momentum. 
At zero temperature the $T=0$ $\beta'=-\beta$ terms contribute,
\begin{align}
\Pi(i\omega,{\bf q}) =&- \frac{N}{4}\sum_{\beta}\int \frac{d^2p}{(2\pi)^2} \frac{\beta}{i\omega + v\beta p +v \beta|{\bf p+q}|} \nonumber\\
&\times\text{Tr}[(1 - \beta{ \sigma}_{\bf p})(1 + \beta {\ \sigma}_{\bf p+q})].
\end{align}
The retarded counterpart of the polarization operator can be obtained through analytical continuation $i\omega\rightarrow\omega+i\eta$. However, since virtual transitions avoid all singular poles, a small imaginary constant $i\eta$ can be disregarded for our purposes,
\begin{equation}
\label{polarization operator}
\Pi(\omega,{\bf q}) = \frac{N}{8\pi^2}\sum_{\beta}\int\limits_{-\infty}^\infty d^2p~ \frac{\beta[\cos(\theta_{\bf p+q}-\theta_{\bf p})-1]}{\omega + v\beta p +v \beta|{\bf p+q}| }.
\end{equation}
Within the zone folding approximation, which ignores any curvature effects on the electronic spectrum arising from the rolling of the graphene sheet, the polarization function for a nanotube is obtained by quantizing the circumferential momenta, $q_y=m/R$, $p_y=n/R$, and replacing the integral with the sum, $R \int dp_y \to \sum_m$. Additionally, to relate the resulting polarization function to the one-dimensional density, the quantity (\ref{polarization operator}) should be multiplied by the factor $2\pi R$. Expressing the cosine function in terms of the momentum component, we obtain:
\begin{widetext}
\begin{eqnarray}
\label{widetext pi}
\Pi_{m}(\omega,q_x) &=& \frac{N}{2\pi}\sum_n\int\limits_{-\infty}^\infty dp_x ~ \Bigl[1-\frac{(p_x+q_x)p_x+(m+n)n/R^2}{\sqrt{(p_x^2+n^2/R^2)}\sqrt{(p_x+q_x)^2+(m+n)^2/R^2}}\Bigr]
\nonumber\\
&&\times\frac{v\sqrt{(p_x^2+m^2/R^2)} + v\sqrt{(p_x+q_x)^2+(m+n)^2/R^2}}{\omega^2 - [v\sqrt{(p_x^2+m^2/R^2)} + v\sqrt{(p_x+q_x)^2+(m+n)^2/R^2}]^2}.
\end{eqnarray}
\end{widetext}
Of interest to us here are the polarization function for $m=0$ and $m=\pm 1$. For $m=0$ and $q_x \ll 1/R$, only the $n=0$ terms should be retained. For example, for $q_x>0$, the integral only extends over the interval $-q_x<p_x<1$ (where the integrand does not depend on $p_x$) and the expression (\ref{polarizationV_0}) is recovered, the subscript in $q_x$ being omitted. (The same result follows for $q_x<0$.)

For $m=1$, because only the static and homogeneous limit is important for our purposes, one can set $\omega=0$ and $q_x=0$ in Eq.~(\ref{widetext pi}). The remaining $p_x$-integral is independent of $R$ and can be (together with the summation over $m$) calculated numerically. This yields
\begin{equation}
\label{polarization one}
\Pi_1(0,0) =
-1.16\frac{N}{\pi v}
\end{equation}
which yields the value of the screened $U_1$ interaction as in Eq.~(\ref{U1}).

Note that the dominant contribution into the polarization function (\ref{polarization one}) comes from the lowest-order virtual transitions, $n=-1$ and $n=0$. Retaining only these contributions, one would obtain the estimate $\Pi_1(0,0) =-\frac{N}{\pi v}$. Another good order-of-magnitude estimate could be obtained from the polarization function for graphene, $\Pi(\omega, q) =-Nq^2/ (16 \sqrt{q^2v^2-\omega^2 })$. In the static limit, $\omega=0$, replacing $q\to 1/R$, and multiplying the result by $2\pi R$, as explained above, one would obtain, $\Pi_1(0,0) =-\frac{\pi N}{8 v}$, which overestimates the exact value (\ref{polarization one}) by only $6\%$.


\begin{thebibliography}{50}

\bibitem{Dresselhaus} R. Saito, G. Dresselhaus, and M. S. Dresselhaus, {\it Physical Properties of Carbon Nanotubes} (Imperial College Press, London, 1998).

\bibitem{WCKDSLZHS} F. Wang, D. J. Cho, B. Kessler, J. Deslippe, P. J. Schuck, S. G. Louie, A. Zettl, T. F. Heinz, and Y. R. Shen, Phys. Rev. Lett. {\bf 99}, 227401 (2007).

\bibitem{MTZRTM} P. May, H. Telg, G. Zhong, J. Robertson, C. Thomsen, and J. Maultzsch, Phys. Rev. B {\bf 82}, 195412 (2010).

\bibitem{Ando} T. Ando, J. Phys. Soc. Jpn. {\bf 66}, 1066 (1997).

\bibitem{SBBL} C. D. Spataru, S. Ismail-Beigi, L. X. Benedict, and S. G. Louie, Phys. Rev. Lett. {\bf 92}, 077402 (2004).

\bibitem{DSPL} J. Deslippe, C. D. Spataru, D. Prendergast, and S. G. Louie, Nano Lett. {\bf 7}, 1626 (2007).

\bibitem{UA} S. Uryu and T. Ando, Phys. Rev. B {\bf 77}, 205407 (2008).

\bibitem{MMRK} E. Malic, J. Maultzsch, S. Reich, and A. Knorr, Phys. Rev. B {\bf 82}, 035433 (2010).

\bibitem{AU} T. Ando and S. Uryu, Phys. Status Solidi C {\bf 6}, 173 (2009).

\bibitem{Loudon} R. Loudon, Am. J. Phys. {\bf27}, 649 (1959).

\bibitem{Andrew1} M. Andrew. Am. J. Phys. {\bf 34}, 1194, (1966).

\bibitem{HR} L. K. Haines, and D. H. Roberts, Am. J. Phys. {\bf 37}, 1145, (1969).

\bibitem{Andrew2} M. Andrew, Am. J. Phys. {\bf 44}, 1064, (1976).

\bibitem{GZ} J. F. Gomes, and A. H. Zimerman, Am. J. Phys. {\bf 48}, 579, (1980).

\bibitem{Moss} R. E. Moss, Am. J. Phys. {\bf 55}, 397, (1987).

\bibitem{KSSMSN} T. Koyama, S. Shimizu, T. Saito, Y. Miyata, H. Shinohara, and A. Nakamura, Phys. Rev. B {\bf 85}, 045428 (2012).

\bibitem{KCM} M. K. Kostov, M. W. Cole, and G. D. Mahan, Phys. Rev. B {\bf 66}, 075407 (2002).

\bibitem{Petersen} T. G. Pedersen, Phys. Rev. B {\bf 67}, 073401 (2003).

\bibitem{MAG} E. G. Mishchenko, A.V. Andreev, and L. I. Glazman, Phys. Rev. Lett. {\bf 87}, 246801 (2001).

\bibitem{DL} I. E. Dzyaloshinskii and A. I. Larkin, Sov. Phys. JETP {\bf 38}, 202 (1974).



\end{thebibliography}
\end{document}